# Polarization selectivity of charge localization induced by 7-fs nearly single-cycle light-field in an organic metal

Y. Kawakami[1], Y. Yoneyama[1], T. Amano[1], H. Itoh[1], K. Yamamoto[2],

Y. Nakamura[3], H. Kishida[3], T. Sasaki[4], S. Ishihara[1], Y. Tanaka[5],

K. Yonemitsu[5], and S. Iwai[1*]

[1]Department of Physics, Tohoku University, Sendai 980-8578, Japan

[2]Department of Applied Physics, Okayama University of Science, Okayama, 700-0005 Japan

[3] Department of Applied Physics, Nagoya University, Nagoya 464-8603 Japan

[4]Institute for Materials Research, Tohoku University, Sendai 980-8577, Japan

[5]Department of Physics, Chuo University, Tokyo 112-8551, Japan

71.27.+a, 75.30.Wx, 78.47.jg

*s-iwai@m.tohoku.ac.jp

Abstract

Polarization selectivity of light-field-induced charge localization was investigated in an organic metal α-(BEDT-TTF)$_2$I$_3$ with a triangular lattice. Dependences of transient reflectivity spectra on polarizations of the 7-fs pump and probe lights indicate that a short-range charge order (CO) is efficiently induced from the metallic phase for the pump polarization perpendicular to the 1010-type CO axis. Numerical solution of a time-dependent Schrödinger equation clarified that the 1010-CO is induced by field-induced re-distribution of charges cooperating with competing inter-site Coulomb repulsions in the triangular lattice.

In the progress of light-induced control of correlated materials [1-6], electronic and/or structural orders such as charge density wave (CDW), superconductivity(SC) and charge ordering (CO) triggered by an optical excitation are focused on and anticipated to proceed toward bidirectional switching of phases and discovery of new nonequilibrium phases[7-14]. They are in contrast to well-understood "melting" of orders [1-3]. Central issue is to clarify mechanisms of such "counter intuitive" responses reflecting respective material characters. For light-induced CDW and SC, enhanced Fermi surface nesting driven by ionic motion[12], dynamic electron-phonon interaction[13], and phononic excitation[5, 14] have been discussed.

Recently, a short-range CO during a short time scale (~40 fs) induced by a nearly single-cycle electric field of 7-fs, 10 MV/cm (schematic illustration in Fig. 1(a)) has been demonstrated in an organic metal α-(BEDT-TTF)$_2$I$_3$ with a triangular lattice [15]. Such transient localization of charge was originally motivated by "dynamical localization", i.e., the reduction of an effective transfer integral ($t_{eff} = \frac{1}{T}\int_0^T t e^{i\frac{e}{\hbar c}\mathbf{r}_{ij}\cdot\mathbf{A}(t')} dt'$) which is the time average of an original transfer integral $t$ multiplied by the Peierls-phase factor ($A$(t): vector potential, $r_{ij}$: position vector for corresponding transfer process, $T$: period of the oscillating field)[16-18]. Forming of such a dressed charge state with the driving field is quite different from conventional photoinduced melting of CO that is triggered by a real excitation of carriers [19, 20] in the sense that a single-cycle field cannot resonate with any charge transfer (CT) process within the pulse duration [21,22]. Related studies on a contribution from Coulomb repulsion are still going on [4, 23-28], although the microscopic mechanism remains unclear. For clarifying the mechanism and discussing its generality, investigation of polarization selectivity (if any) of such a dressed charge state is very important to disentangle the network of transfer-integral/intersite-Coulomb-repulsion effects. Given that Coulomb repulsion with the characteristic low-symmetric molecular backbone is the origin of the metal-ferroelectric CO phase transition[29-32], it seems essential for the transient charge localization.

The triangular lattice structure of α-(BEDT-TTF)$_2$I$_3$ [29] with crystallographically non-equivalent sites A, A' B, and C linked by $b_1(b_1')$-$b_4(b_4')$, $a_1(a_1')$, $a_2$ and $a_3$ bonds in the CO phase (below $T_{CO}$=135 K) is shown in Fig. 1(b)[30, 31]. A 1010-type CO along $a$ ($a_2$ and $a_3$) bonds results in charge rich/poor zig-zag stripes A-B-A-(rich;black) and A'-C-A'-(poor;white). Above $T_{CO}$ (metallic phase), the sites A and A', the $b_1$-$b_4$ and $b_1'$-$b_4'$ bonds and the $a_1$ and $a_1'$ bonds are respectively equivalent as shown in Fig. 1(c), although a charge imbalance between sites B and C remains even in the metallic phase because of the low symmetric (triclinic $P\bar{1}$) lattice structure [33, 34].

In this study, we have investigated in the metallic phase at 138 K the dependence of the transient reflectivity on polarizations of both 7-fs pump- and probe- pulses (Fig. 1(a)). In contrast to the intuitive expectation, a short range CO along the $a_2$ and $a_3$ bonds is induced efficiently for the pump with polarization perpendicular to the 1010-CO axis.

Transient reflectivity measurements (Fig. 1(a)) were performed by using 7-fs pulses. The central photon energy of the pump light is ~0.8 eV and the probe range covers 0.58–0.95 eV. The method for generating the 7-fs pulse has been reported in the previous paper[15]. In this study, polarizations of both pump and probe pulses were controlled by a respective pair of wire-grid $CaF_2$ polarizers as shown in Fig. 1(a). The light intensity after the wire-grid polarizers is shown as a function of angles $\theta_2$ and $\theta_1$. The pulse width derived from the autocorrelation (Fig. 4(a)) was 7 fs (, which corresponds to 1.5 optical cycles) at the sample position after the wire-grid polarizers and the window of the cryostat. In the transient reflectivity measurements using the 7-fs pulse, the instantaneous electric field on the sample surface (excitation diameter of 200 μm) was evaluated as $E_{peak} = 9.8 \times 10^6 \mathrm{(V/cm)}$ for an excitation intensity $I_{ex}$ of 0.9 mJ/cm². Polarized optical reflection experiments were conducted on a single crystal of α-(BEDT-TTF)$_2$I$_3$ (of size 0.5 × 0.7 × 0.2 mm for crystalline axes *a*, *b*, *c*, respectively [33, 34].) grown by electro-crystallization [29].

The solid lines in Figs. 2 (a) and 2(b) show steady state reflectivity (*R*) spectra measured for various polarizations at 138 K. The spectral differences between 138 K (metal) and 50 K (insulator) $(\Delta R/R)_{MI} = \{R(50K) - R(138K)\}/R(138K)$ are shown for various polarizations by the lines in Fig. 2(c). Here, $\theta$ is defined as the angle between the *b* axis and the light polarization (Fig. 1(a)). Figure 2(c) clearly indicates that *R* markedly increases upon the metal to CO transition for $\theta = 0°$ (//*b*) at > 0.6 eV (the red solid line), whereas change in *R* is hardly detected for $\theta = 90°$ (//*a*) at the same spectral range (the black two-dotted line). It is noteworthy that the metal to CO transition is sensitive to the change in $R$ for $\theta=0°$ (//$b$) in this spectral range. Such polarization dependence of $(\Delta R/R)_{MI}$ is consistent with the theoretical calculation of the optical conductivity based on the exact diagonalization in the frame work of an extended Hubbard model[31]. Thus, the polarization dependence of $(\Delta R/R)_{MI}$ spectrum at the spectral range >0.6 eV is recognized as the fingerprint of the CO caused by charge motion mainly along the $b$ bonds.

The red-filled circles and the blue squares in Fig. 2(c) respectively show transient reflectivity ($\Delta R/R$) spectra for $\theta_{pr}$ (angle between the polarization of the probe pulse and the $b$ axis as shown in Fig. 1(a))=0° (//$b$) and =90° (//$a$) at the delay time ($t_d$) of 30 fs after the excitation at 138 K. Here, $\theta_{pu}$ (angle between the polarization of the pump pulse and the $b$ axis as shown in Fig. 1(a))=0° (//$b$) for both spectra. The $\Delta R/R$ spectrum for $\theta_{pr}=0°$ (//$b$) shows a large (>25 %) increase of $R$ at 0.67 eV, although the $\Delta R/R$ for $\theta_{pr}=90°$ (//$a$) is small (~1%). Figure 3(a) shows the $\Delta R/R$ (0.65 eV) as a function of $\theta_{pr}$ at $t_d=30$ fs. It takes a maximum at $\theta_{pr}=0°$. Such polarization dependence is analogous to that of the $(\Delta R/R)_{IM}$ at 0.65 eV (shown by the dashed line in Fig. 3(a)), which is the finger print of the CO. The photoinduced M-I change has been previously discussed on the basis of the $\Delta R/R$ detected only for $\theta_{pr}=0°$ (//b) [15]. Here, we confirmed that the $\theta_{pr}$ dependence of $\Delta R/R$ at $t_d$=30 fs is analogous to that of the CO.

The time evolutions of the $\Delta R/R$ (0.65 eV) for various $\theta_{pu}$ are shown in Fig. 4 ($\theta_{pr}=0°$ (//b)). The $\Delta R/R>0$ reflecting the CO survives only about 40 fs indicates that the induced CO is neither a long-range order nor a local excited state[35] which is structurally stabilized, because it requires the timescale of intermolecular vibrations (> 100 fs) for stabilization. So the short-lived CO is of short-range without intermolecular structural stabilization. The oscillation with a period of 20 fs has been attributed to an intermolecular charge oscillation reflecting a CO gap, and $\Delta R/R<0$ after $t_d=40$ fs to an increase of electron temperature[15]. Here, the spectrum of the 7 fs broad-band pulse covering 0.58–0.95 eV is set on the higher energy side of the main electronic band peaked at 0.1-0.2 eV to avoid resonance. Moreover, because the pulse width of the nearly single-cycle 7 fs pulse is shorter than or equal to the time scale of CT processes, the transient state during the pulse width should be considered as a dressed charge state with the oscillating field [21, 22].

Figure 3(b) shows that the $\theta_{pu}$ dependence of $\Delta R/R$ ($t_d=30$ fs) at 138 K for various $\theta_{pr}$ (0°, 23°, -23°, 90°). An efficient charge localization to the short-range CO was observed for $\theta_{pu}=0°$ (//*b*) for all $\theta_{pr}$, whereas it hardly occurred for $\theta_{pu}=90°$ (//a). In fact, $\Delta R/R$ for $\theta_{pu}=0°$ is >10 times larger than that for $\theta_{pu}=90°$. Such drastic change of $\Delta R/R$ depending on $\theta_{pu}$ cannot be attributed to the dependence of the absorption coefficient α on $\theta$, i.e., the difference of α is smaller than twice for //a and //b (~10000 cm$^{-1}$ for //a, 20000

cm$^{-1}$ for //b in the spectral range covered by the 7 fs pump pulse). Therefore, the result in Fig. 3(b) is not affected so much by the polarization dependence of α.

A plausible reason why the CO is efficiently induced for $\theta_{pu}$ =0⁰ (//b) is that the reduction of $t_{eff}$ in the $b_1$ and $b_2$ bonds during the single-cycle pulse triggers the CO in the timescale of 40 fs after the excitation. The reduction of $t$ (~6%) has been evaluated in itself as a red shift of plasma-like reflectivity edge in quasi-1-D (TMTTF)$_2$AsF$_6$[27]. However, the drastic charge localization (as in α-(BEDT-TTF)$_2$I$_3$) was not detected in the 1-D system, indicating that the 2-D triangular structure is crucial. To consider a microscopic mechanism, we performed a numerical calculation of the strong light-field effect using a time-dependent Schrödinger equation in the frame work of the 1/4-filled extended Hubbard model[36, 37].

Here, we employ the two-dimensional, 1/4-filled extended Hubbard model for the metallic phase,

$$H = \sum_{\langle i,j \rangle \sigma} t_{ij}\left(c^+_{i,\sigma}c_{j,\sigma} + c^+_{j,\sigma}c_{i,\sigma}\right) + U\sum_i n_{i\uparrow}n_{i\downarrow} + \sum_{\langle i,j \rangle} V_{ij}n_i n_j$$

, where $c_{i\sigma}$ is the annihilation operator of a hole on site $i$ with spin $\sigma$, $n_{i,\sigma} = c^+_{i\sigma}c_{i\sigma}$, and $n_i = \sum_\sigma n_{i,\sigma}$ [36]. This model has on-site Coulomb repulsion ($U$), nearest-neighbor repulsions ($V_{ij}$) and transfer integrals between sites $i$ and $j$ ($t_{ij}$). We used the exact diagonalization method for the 16-site system with periodic boundary conditions, and set $t_{ij}$, which were evaluated by the extended Hückel calculation using the X-ray structural analysis at the metallic phase[30]. Photoexcitation is introduced through the Peierls phase. By numerically solving the time-dependent Schrödinger equation during and after the excitation of a single-cycle pulse with central frequency $\hbar\omega$ = 0.8 eV, we calculated the change in correlation functions: the spatially and temporally averaged double occupancy $\langle\langle n_{i\uparrow}n_{i\downarrow}\rangle\rangle$, and averaged nearest-neighbor density-density correlations $\langle\langle n_i n_j\rangle\rangle_{a2,a3}$ for the $a_2$ and $a_3$ bonds and $\langle\langle n_i n_j\rangle\rangle_b$ for the $b$ bonds, as indexes of the short-range 1010-CO along the $a_2$ and $a_3$ bonds[38]. Temporal average was calculated for $5T < t_d < 10T$. The calculation is performed on finite-size systems, so that the symmetry cannot be broken spontaneously and no long-range CO is produced. However, our calculation is useful for discussing the generation of a short-range CO.

Figure 5 shows (a) $\langle\langle n_{i\uparrow}n_{i\downarrow}\rangle\rangle$, (b) $\langle\langle n_i n_j\rangle\rangle_{a2,a3}$ and (c) $\langle\langle n_i n_j\rangle\rangle_b$ as functions of $eaF/\hbar\omega$ ($a$: component parallel to the $b$ axis of the molecular spacing, $F$: field amplitude) for $U$=0.8 eV, intersite Coulomb repulsions $V_1$(along all $b$ bonds)=0.3 eV, $V_2$ (along all $a$ bonds) =0.35 eV. As shown in Fig. 5(a), $\langle\langle n_{i\uparrow}n_{i\downarrow}\rangle\rangle$ decreases with increasing $eaF/\hbar\omega$ for <0.4, which means that holes with opposite spins avoid being on the same site more strongly as if $U$ were increased relative to the transfer integrals transiently after excitation. If we reproduce the decrease of $\langle\langle n_{i\uparrow}n_{i\downarrow}\rangle\rangle$ (14 % for $eaF/\hbar\omega$ =0.37) in the ground state, we need an increase of $U$ by 6.5%. On the other hand, the decrease of $\langle\langle n_i n_j\rangle\rangle_{a2,a3}$ (15 % for $eaF/\hbar\omega$ =0.37) (Fig. 5(b)) means that holes avoid neighboring along the $a_2$ and $a_3$ bonds more strongly as if the $V_{ij}$ along the $a_2$ and $a_3$ bonds were increased relative to the transfer integrals transiently after photoexcitation.

Figure 5(d) shows the $\theta_{pu}$ dependence of the averaged nearest-neighbor density-density correlation for the $a_2$ and $a_3$ bonds $\langle\langle n_i n_j \rangle\rangle_{a2,a3}$ reflecting the 1010-type CO along the $a_2$ and $a_3$ bonds for $eaF/\hbar\omega$ =0 (black crosses), 0.14 (blue rhombuses), 0.20 (green triangles), 0.27 (red squares) ($U$= 0.8 eV, $V_1$=0.3 eV, $V_2$=0.35 eV). A reduction of $\langle\langle n_i n_j \rangle\rangle_{a2,a3}$ is efficient at $\theta_{pu}$ =0~10⁰ which is almost parallel to the $b$ axis for any $eaF/\hbar\omega$. Such polarization dependence of $\langle\langle n_i n_j \rangle\rangle_{a2,a3}$ is consistent with that of $\Delta R/R$ (Fig. 3(b)) reflecting the construction efficiency of the CO within the accuracy of $\theta_{pu}$.

Figures 5(e) and 5(f) show the hole densities on sites A(averaged over sites A and A'), B and C ($n_A$, $n_B$ and $n_C$) as functions of $eaF/\hbar\omega$. The charges on sites A and A' are directly related to the 1010-CO (Fig. 1(b)). As shown in Fig. 5(f), $n_A$ is larger than 0.5 which is expected in the 1/4-filled band at zero field reflecting the relation $|t_{b2}|>|t_{b1}|$ ($t_{b1}$, and $t_{b2}$ are the $t$ in the $b_1$ and $b_2$ bonds). It is noteworthy that $n_A$ approaches to 0.5 with increasing $eaF/\hbar\omega$. Considering that the averaged hole density 0.5 on sites A and A' is ideal for the CO to be driven by $V_2$ along the $a_2$ and $a_3$ bonds, the field-induced decrease of $n_A$ from 0.54 to 0.5 is important for inducing the 1010-CO along the $a_2$ and $a_3$ bonds. In this mechanism, the repulsion $V_2$ slightly larger than the repulsion $V_1$ is indispensable [36].

In Fig. 5(e), the charge imbalance between sites B and C at zero field remains even in the metallic phase (Fig. 1(c)), because of a large transfer integral on the $b_2$ bond. Such imbalance between $n_B$ (0.69) and $n_C$ (0.23) is weakened with increasing $eaF/\hbar\omega$. This result indicates that the charge imbalance on sites B-C in the short-range 1010-CO is weaker than that in the CO phase below $T_{CO}$. Thus, the comparable Coulomb repulsions $V_2$ and $V_1$ in the characteristic low-symmetric molecular backbone are essential for the transient charge localization. That is supported by the fact that any dramatic localization has not been detected in a quasi 1-D organic metal (TMTTF)$_2$AsF$_6$ [27]. The correlation related mechanism of the transient CO in the present study is conceptionally different from the mechanisms for light-induced CDW[12, 13] and SC[5, 14]. It opens a new strategy for light-induced control of correlated electron materials.

In summary, the dependence of the transient reflectivity spectra on polarizations of 7-fs pump and probe lights were investigated for clarifying the controllability and the mechanism for transiently inducing the short-range CO in an organic metal α-(BEDT-TTF)$_2$I$_3$. Efficient induction of the short-range CO for the pump polarization perpendicular to the 1010-CO axis is realized by re-distribution of charges through the $b$ bonds and competing inter-site Coulomb interactions along the $a$ and $b$ bonds in the triangular lattice.

Mol. Cryst. Liq. Cryst. 138, 393 (1986).

[35] M. Servol, N. Moisan, E. Collet, H. Cailleau, W. Kaszub, L. Toupet, D. Boschetto, T. Ishikawa, A. Moreac, S. Koshihara, M. Maesato, M. Uruichi, X. Shao, Y. Nakano, H. Yamochi, G. Saito, and M. Lorenc, Phys. Rev. B, 92, 024304(2015).

[36] K. Yonemitsu, *J. Phys. Soc. Jpn.*, **86**, 024711(2017).

[37] In [36], calculations were made on an equilateral triangular lattice. On the other hand, a real structure of this compound obtained by the X-ray analysis[30] was taken into account in this study.

[38] Dynamical localization can be described as a reduction of an effective transfer integral. However, we never use effective transfer integrals, when we solve the time-dependent Schrödinger equation. The transfer integrals we use are directly derived from the extended Hückel calculation based on the X-ray structural data in the metallic phase[30]. In general, effective transfer integrals are merely a way to understand the numerical results after averaging over the time scale of the field oscillation period: they are obtained by averaging the transfer integrals multiplied by the time-dependent Peierls-phase factor, which is always a complex number of modulus one. Effective transfer integrals are a useful concept and a key element of dynamical localization. However, they are not sufficient to understand the numerical results. Thus, we calculate the time averages of different correlation functions, double occupancy $\langle\langle n_{i\uparrow} n_{i\downarrow} \rangle\rangle$ and nearest-neighbor correlation functions $\langle\langle n_i n_j \rangle\rangle$, which are modified by the

field in different manners.

The double occupancy $\langle\langle n_{i\uparrow}n_{i\downarrow}\rangle\rangle$ is a measure of localization, and nearest-neighbor correlation functions $\langle\langle n_i n_j\rangle\rangle$ are measures of how neighboring electrons avoid each other and characterize the short-range CO. For the 1010 CO, $\langle\langle n_i n_j\rangle\rangle_{a2,a3}$ for the $a_2$, $a_3$ bonds should be reduced as compensation for increased $\langle\langle n_i n_j\rangle\rangle_b$ for the *b* bonds. The increased $\langle\langle n_i n_j\rangle\rangle_b$ for the *b* bonds are compatible with charge-rich stripes along A(') and B sites. Because the charge disproportionation can be induced by both Coulomb repulsion and kinetic energy on such a low symmetric lattice structure, observation of these correlation functions is essential.

# Figure captions

Figure 1

(a) A light-induced short-range CO and its detection by pump-probe reflectivity measurement. The light intensity after a pair of wire-grid CaF$_2$ polarizers is also shown as a function of angles $\theta_1$ and $\theta_2$. (b)(c) Lattice structure of α-(BEDT-TTF)$_2$I$_3$ in CO phase (b) and that in metallic phase (c). Sites(=BEDT-TTF molecules) A, A', B and C are crystallographically non-equivalent. $b_1(b_1')$-$b_4(b_4')$, $a_1(a_1')$, $a_2$ and $a_3$ are the bonds between those sites.

Figure 2

(a)(b) Reflectivity ($R$) spectra (0.08-0.8 eV (a), 0.4-0.8 eV(b)) measured for various polarizations ($\theta$ =0º, 20º, 40º, 60º, 90º) at 138 K. (c) Transient reflectivity ($\Delta R/R$) spectra at the delay time ($t_d$) of 30 fs are shown for $\theta_{pr}$ =0º (//b) (red-filled circles) and $\theta_{pr}$ =90º (//a) (blue squares) for $\theta_{pu}$ =0º (//b). The excitation intensity $I_{ex}$ was 0.9 mJ/cm² (9.8 MV/cm). The spectral differences between 138 K (metal) and 50 K (insulator) $(\Delta R/R)_{MI} = \{R(50K) - R(138K)\}/R(138K)$ are also shown for $\theta$ =0º, 20º, 40º, 60º, 90º.

Figure 3

(a) Transient reflectivity ($\Delta R/R$) measured at 0.65 eV, $t_d$ =30 fs as a function of $\theta_{pr}$ ($\theta_{pu}$ =0º). The dashed line shows a polarization dependence of

$(\Delta R/R)_{MI} = \{R(50K) - R(138K)\}/R(138K)$ at 0.65 eV (x0.3). (b) $\Delta R/R$ at 0.65 eV, $t_d$ =30 fs as a function of $\theta_{pu}$ for $\theta_{pr}$ =0°(red filled circles), 23°(green rectangles), -23°(orange triangles), 90°(blue triangles).

Figure 4

(a) Autocorrelation function of the 7-fs pulse measured at the sample position after the wire-grid polarizers and the window of the cryostat. (b) Time evolutions of $\Delta R/R$ measured at 0.65 eV for $\theta_{pu}$ =0°(black solid line), 23°(blue dashed line), 40°(green dotted line), 60°(orange dashed-dotted line) and 90°(red dashed two-dotted line) . $\theta_{pr}$ was set at 0° for all $\theta_{pu}$.

Figure 5

(a)-(c) Field induced changes in spatially and temporally averaged correlation functions as indexes of the short-range CO. Double occupancy (a) $\langle\langle n_{i\uparrow} n_{i\downarrow} \rangle\rangle$. Nearest-neighbor density-density correlations (b) $\langle\langle n_i n_j \rangle\rangle_{a2,a3}$ for the $a_2$ and $a_3$ bonds and (c) $\langle\langle n_i n_j \rangle\rangle_b$ for the $b$ bonds. (d) $\theta_{pu}$ dependence of the averaged nearest-neighbor density-density correlation for the $a_2$ and $a_3$ bonds $\langle\langle n_i n_j \rangle\rangle_{a2,a3}$ for $eaF/\hbar\omega$ =0 (black crosses), 0.14 (blue rhombuses), 0.20 (green triangles), and 0.27 (red squares). (e)(f) Hole densities on sites A(averaged over sites A and A'), B and C ($n_A$, $n_B$ and $n_C$) as functions of $eaF/\hbar\omega$.

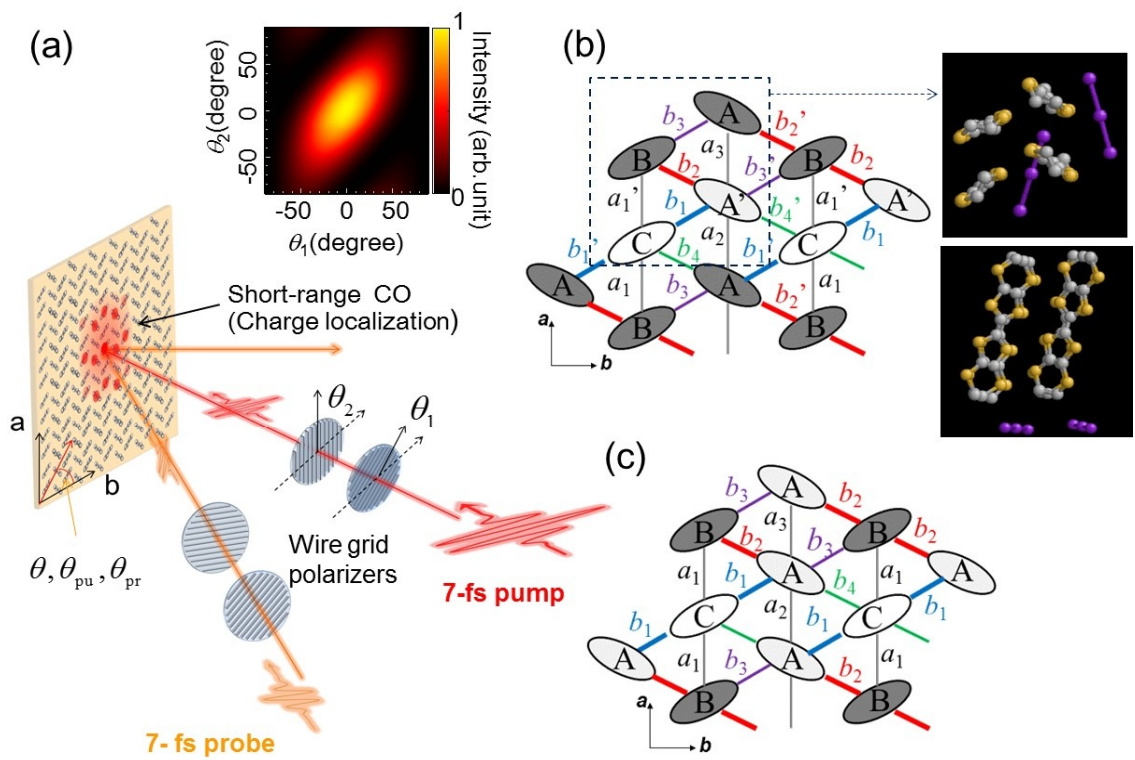

Kawakami et al., Fig. 1

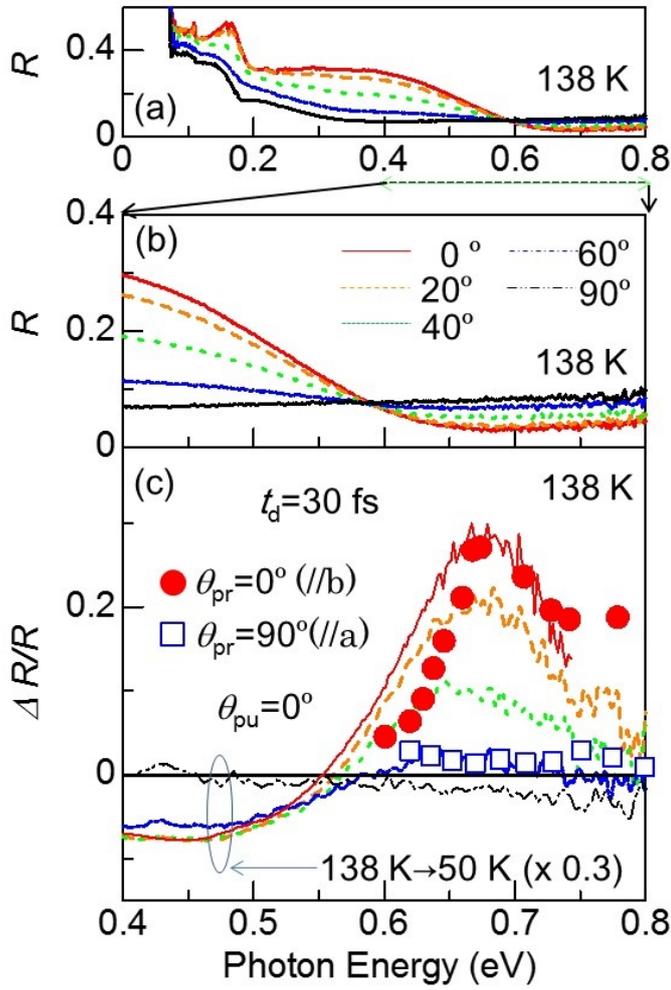

Kawakami et al., Fig. 2

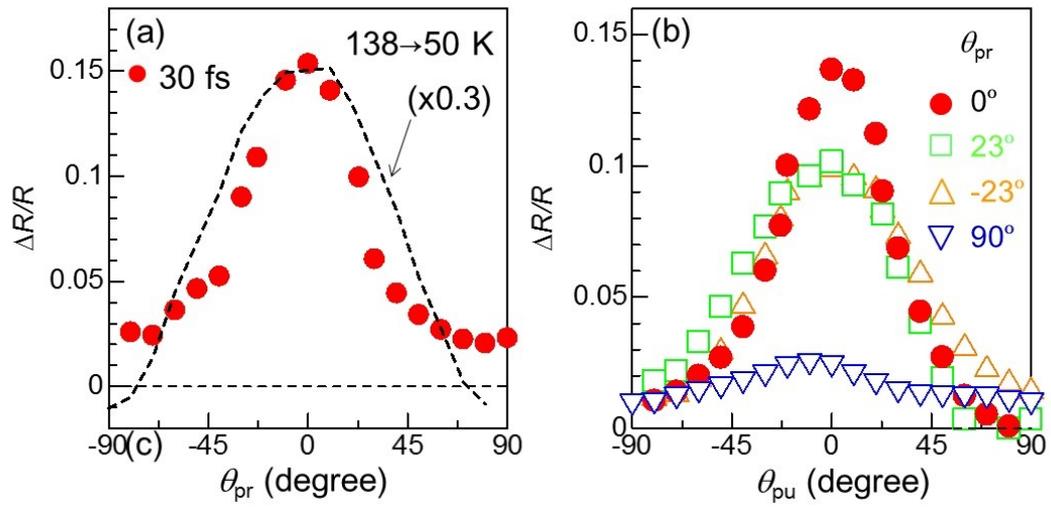

Kawakami et al., Fig.3

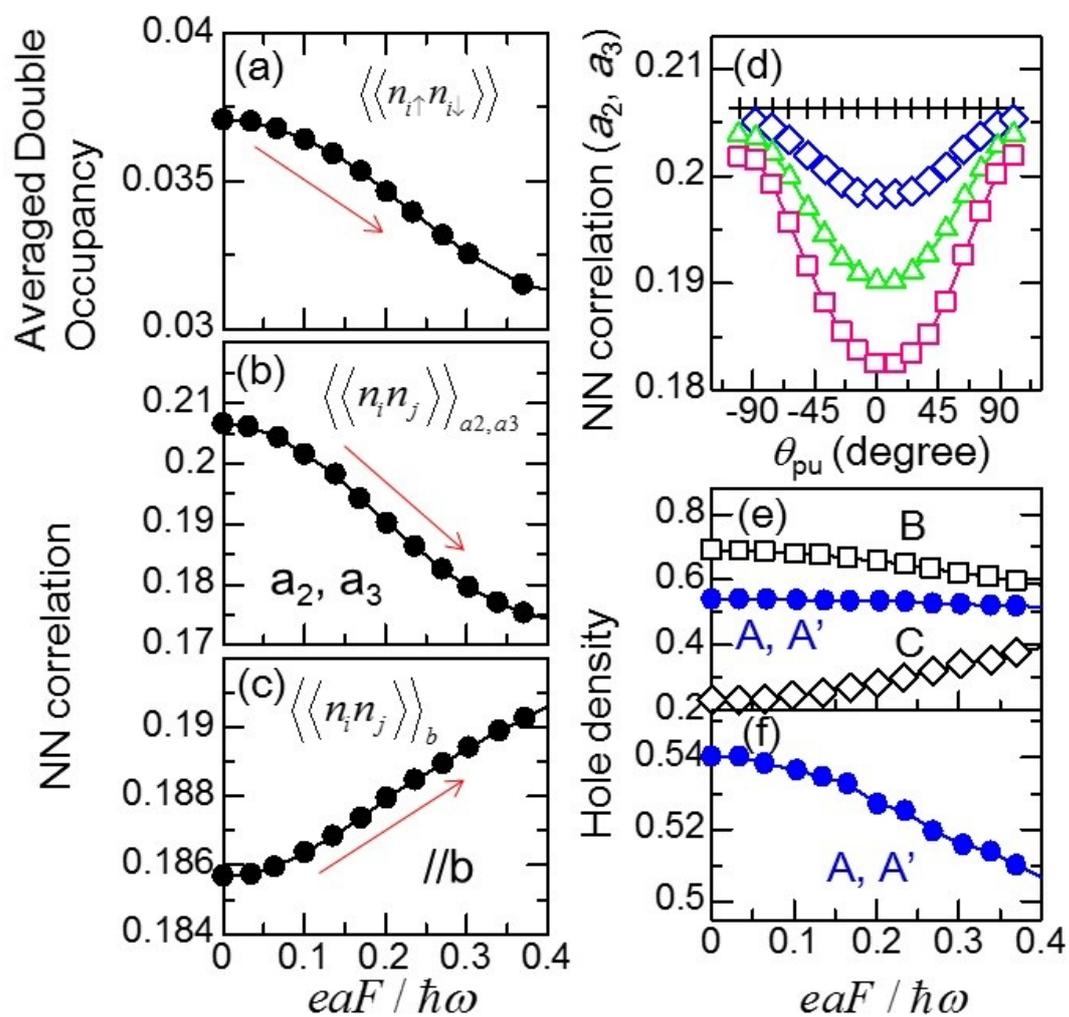

Kawakami et al., Fig.5